\documentclass[12pt,preprint]{aastex}
\usepackage{bm}
\usepackage{threeparttable}
\usepackage{amsmath,amssymb}
\usepackage{textcomp}
\usepackage{booktabs}
\usepackage{lscape}
\makeatletter
\newcommand{\tblcaption}[1]{\def\@captype{table}\caption{#1}}
\makeatother
\usepackage{color}
\usepackage{comment}

\title{OGLE-2015-BLG-1649Lb: A gas giant planet around a low-mass dwarf}

\author{M. Nagakane\altaffilmark{1}, Chien-Hsiu Lee\altaffilmark{2}, N. Koshimoto\altaffilmark{3,4}, D. Suzuki\altaffilmark{5}, A. Udalski\altaffilmark{6}, J. P. Beaulieu\altaffilmark{7}, T. Sumi\altaffilmark{1}, D. P. Bennett\altaffilmark{8,9}, I. A. Bond\altaffilmark{10}, N. Rattenbury\altaffilmark{11}, E. Bachelet\altaffilmark{12}, M. Dominik\altaffilmark{13}}

\and

\author{F. Abe\altaffilmark{14}, R. K. Barry\altaffilmark{8}, A. Bhattacharya\altaffilmark{8,9}, M. Donachie\altaffilmark{11}, H. Fujii\altaffilmark{14}, A. Fukui\altaffilmark{3, 15}, Y. Hirao\altaffilmark{1}, Y. Itow\altaffilmark{14}, Y. Kamei\altaffilmark{14}, I. Kondo\altaffilmark{1}, M. C. A. Li\altaffilmark{11}, Y. Matsubara\altaffilmark{14}, T. Matsuo\altaffilmark{1}, S. Miyazaki\altaffilmark{1} Y. Muraki\altaffilmark{14}, C. Ranc\altaffilmark{8}, H. Shibai\altaffilmark{1}, H. Suematsu\altaffilmark{1}, D. J. Sullivan\altaffilmark{16}, P. J. Tristram\altaffilmark{17}, T. Yamakawa\altaffilmark{14}, A. Yonehara\altaffilmark{18}\\ (The MOA Collaboration)}

\author{P. Mr{\'o}z\altaffilmark{6}, R. Poleski\altaffilmark{19,6}, J. Skowron\altaffilmark{6}, M. K. Szyma{\'n}ski\altaffilmark{6}, I. Soszy{\'n}ski\altaffilmark{6}, P. Pietrukowicz\altaffilmark{6}, S. Koz{\l}owski\altaffilmark{6}, K. Ulaczyk\altaffilmark{20,6} \\(The OGLE Collaboration)}

\author{D. M. Bramich\altaffilmark{21,22,23}, A. Cassan\altaffilmark{24}, R. Figuera Jaimes\altaffilmark{25}, K. Horne\altaffilmark{26}, M. Hundertmark\altaffilmark{27}, S. Mao\altaffilmark{28,29}, J. Menzies\altaffilmark{30}, R. Schmidt\altaffilmark{27}, C. Snodgrass\altaffilmark{31,32}, I. A. Steele\altaffilmark{33}, R. Street\altaffilmark{12}, Y. Tsapras\altaffilmark{27}, J. Wambsganss\altaffilmark{27,34} \\(The RoboNet Collaboration)}

\author{U. G. J{\o}rgensen\altaffilmark{35}, V. Bozza\altaffilmark{36,37}, P.Long{\~a}\altaffilmark{38}, N. Peixinho\altaffilmark{38,39}, J. Skottfelt\altaffilmark{31}, J. Southworth\altaffilmark{40}, M.I. Andersen\altaffilmark{41}, M. J. Burgdorf\altaffilmark{42}, G. D'Ago\altaffilmark{43}, D. F. Evans\altaffilmark{40}, T. C. Hinse\altaffilmark{44}, H. Korhonen\altaffilmark{41}, M. Rabus\altaffilmark{45,46}, S. Rahvar\altaffilmark{47} \\(The MiNDSTEp Collaboration)}

\altaffiltext{1}{Department of Earth and Space Science, Graduate School of Science, Osaka University, 1-1 Machikaneyama, Toyonaka, Osaka 560-0043, Japan}
\altaffiltext{2}{650 N Aohoku Pl. Hilo, HI 96720, USA}
\altaffiltext{3}{Department of Astronomy, Graduate School of Science, The University of Tokyo, 7-3-1 Hongo, Bunkyo-ku, Tokyo 113-0033, Japan}
\altaffiltext{4}{National Astronomical Observatory of Japan, 2-21-1 Osawa, Mitaka, Tokyo 181-8588, Japan}
\altaffiltext{5}{Institute of Space and Astronautical Science, Japan Aerospace Exploration Agency, 3-1-1 Yoshinodai, Chuo, Sagamihara, Kanagawa 252-5210, Japan}
\altaffiltext{6}{Astronomical Observatory, University of Warsaw, Al.~Ujazdowskie~4,00-478~Warszawa, Poland}
\altaffiltext{7}{Sorbonne Universit{\'e}s, CNRS, UPMC Univ Paris 06, UMR 7095, Institut d'Astrophysique de Paris, F-75014, Paris, France}
\altaffiltext{8}{Laboratory for Exoplanets and Stellar Astrophysics, NASA/Goddard Space Flight Center, Greenbelt, MD 20771, USA}
\altaffiltext{9}{Department of Astronomy, University of Maryland, College Park, MD 20742, USA}
\altaffiltext{10}{Institute of Information and Mathematical Sciences, Massey University, Private Bag 102-904, North Shore Mail Centre, Auckland, New Zealand}
\altaffiltext{11}{Department of Physics, University of Auckland, Private Bag 92019, Auckland, New Zealand}
\altaffiltext{12}{Las Cumbres Observatory Global Telescope Network, 6740 Cortona Drive, suite 102, Goleta, CA 93117, USA}
\altaffiltext{13}{Centre for Exoplanet Science, SUPA, School of Physics \& Astronomy, University of St Andrews, North Haugh, St Andrews KY16 9SS, UK}
\altaffiltext{14}{Institute for Space-Earth Environmental Research, Nagoya University, Nagoya, 464-8601, Japan}
\altaffiltext{15}{Instituto de Astrof\'isica de Canarias, V\'ia L\'actea s/n, E-38205 La Laguna, Tenerife, Spain}
\altaffiltext{16}{School of Chemical and Physical Sciences, Victoria University, Wellington, New Zealand}
\altaffiltext{17}{University of Canterbury Mt. John Observatory, P.O. Box 56, Lake Tekapo 8770, New Zealand}
\altaffiltext{18}{Department of Physics, Faculty of Science, Kyoto Sangyo University, Kyoto 603-8555, Japan}
\altaffiltext{19}{Department of Astronomy, Ohio State University, 140 W. 18th Ave., Columbus, OH~43210,~USA}
\altaffiltext{20}{Department of Physics, University of Warwick, Gibbet Hill Road, Coventry, CV4~7AL,~UK}
\altaffiltext{21}{Center for Space Science, NYUAD Institute, New York University Abu Dhabi, PO Box 129188, Saadiyat Island, Abu Dhabi, UAE}
\altaffiltext{22}{Division of Science, New York University Abu Dhabi, PO Box 129188, Saadiyat Island, Abu Dhabi, UAE}
\altaffiltext{23}{Division of Engineering, New York University Abu Dhabi, PO Box 129188, Saadiyat Island, Abu Dhabi, UAE}
\altaffiltext{24}{Institut d'Astrophysique de Paris, Sorbonne Universit{\'e}, CNRS, UMR 7095, 98 bis bd Arago, 75014 Paris, France}
\altaffiltext{25}{Facultad de Ingenieria y Tecnologia Universidad San Sebastian, General Lagos 1163, Valdivia 5110693, Chile}
\altaffiltext{26}{SUPA, University of St Andrews, School of Physics \& Astronomy, North Haugh, St Andrews, KY16 9SS, UK}
\altaffiltext{27}{Zentrum f{\"u}r Astronomie der Universit{\"a}t Heidelberg, Astronomisches Rechen-Institut, M{\"o}nchhofstr. 12-14, 69120 Heidelberg, Germany}
\altaffiltext{28}{Physics Department and Tsinghua Centre for Astrophysics, Tsinghua University, Beijing 100084, China}
\altaffiltext{29}{National Astronomical Observatories, Chinese Academy of Sciences, 20A Datun Road, Chaoyang District, Beijing 100012, China}
\altaffiltext{30}{South African Astronomical Observatory, PO Box 9, Observatory 7935, South Africa}
\altaffiltext{31}{Planetary and Space Sciences, Department of Physical Sciences, The Open University, Milton Keynes, MK7 6AA, UK}
\altaffiltext{32}{Institute for Astronomy, University of Edinburgh, Royal Observatory, Edinburgh EH9 3HJ, UK}
\altaffiltext{33}{Astrophysics Research Institute, Liverpool John Moores University, Liverpool CH41 1LD, UK}
\altaffiltext{34}{International Space Science Institute (ISSI), Haller-strasse 6, 3012 Bern, Switzerland}
\altaffiltext{35}{Niels Bohr Institute \& Centre for Star and Planet Formation, University of Copenhagen, {\O}ster Voldgade 5, 1350 Copenhagen, Denmark}
\altaffiltext{36}{Dipartimento di Fisica "E.R. Caianiello", Universit{\'a} di Salerno, Via Giovanni Paolo II 132, 84084, Fisciano, Italy}
\altaffiltext{37}{Istituto Nazionale di Fisica Nucleare, Sezione di Napoli, Napoli, Italy}
\altaffiltext{38}{Centro de Astronom{\'i}a (CITEVA), Universidad de Antofagasta, Avda.\ U.\ de Antofagasta 02800, Antofagasta, Chile}
\altaffiltext{39}{CITEUC -- Center for Earth and Space Research of the University of Coimbra, Geophysical and Astronomical Observatory, R. Observat{\'o}rio s/n, 3040-004 Coimbra, Portugal}
\altaffiltext{40}{Astrophysics Group, Keele University, Staffordshire, ST5 5BG, UK}
\altaffiltext{41}{Niels Bohr Institute, Juliane Maries vej 30, 2100 Copenhagen, Denmark}
\altaffiltext{42}{Universit{\"a}t Hamburg, Faculty of Mathematics, Informatics and Natural Sciences, Department of Earth Sciences, Meteorological Institute, Bundesstra\ss{}e 55, 20146 Hamburg, Germany}
\altaffiltext{43}{Centro de Astroingenier{\'i}a, Facultad de F\'isica, Pontificia Universidad Cat\'olica de Chile, Av. Vicu\~na Mackenna 4860, Macul 7820436, Santiago, Chile}
\altaffiltext{44}{Chungnam National University, Department of Astronomy and Space Science, 34134 Daejeon, Republic of Korea}
\altaffiltext{45}{Las Cumbres Observatory Global Telescope, 6740 Cortona Dr., Suite 102, Goleta, CA 93111, USA}
\altaffiltext{46}{Department of Physics, University of California, Santa Barbara, CA 93106-9530, USA}
\altaffiltext{47}{Department of Physics, Sharif University of Technology, PO Box 11155-9161 Tehran, Iran}

\keywords{gravitational lensing: micro: planetary system}

\newpage
\begin{document}

\begin{abstract}

We report the discovery of an exoplanet  from the analysis of the gravitational microlensing event OGLE-2015-BLG-1649 that challenges the core accretion model of planet formation and appears to support the disk instability model.
The planet/host-star mass ratio is $q =7.2 \times 10^{-3}$ and the projected separation normalized to the angular Einstein radius is $s = 0.9$. 
We conducted high-resolution follow-up observations using the IRCS camera on the Subaru telescope and are able to place an upper limit on the lens flux. 
From these measurements we are able to exclude all host stars greater than or equal in mass to a G-type dwarf.
We conducted a Bayesian analysis with these new flux constraints included as priors resulting in estimates of the masses of the host star and planet. 
These are $M_{L} = 0.34 \pm 0.19 ~ M_{\odot}$ and $M_{p} = 2.5^{+1.5}_{-1.4} ~ M_{Jup}$, respectively.
The distance to the system is $D_{L} = 4.23^{+1.51}_{-1.64} ~$kpc.
The projected star-planet separation is $a_{\perp} = 2.07^{+0.65}_{-0.77} ~$AU.
The estimated relative lens-source proper motion, $\sim 7.1 ~ {\rm mas/yr}$, is fairly high and thus the lens can be better constrained if additional follow-up observations are conducted several years after the event. 

\end{abstract}

\maketitle
\newpage

\section{Introduction}
\label{sec-intro}

More than 4000 exoplanets have been confirmed since the discovery of the first exoplanet orbiting a main-sequence star, 51 Pegasi b, in 1995 \citep{1995Natur.378..355M}.
A great portion of these planets were discovered using the radial-velocity \citep{2006ApJ...646..505B} and the transit methods \citep{2011ApJ...736...19B}.
While these methods are most sensitive to giant planets in close orbits, the Kepler mission \citep{2010Sci...327..977B}, using the transit technique, demonstrated sensitivity to planets as small as Mercury with semi-major axes of about 1AU. 
A consequence of the limited sensitivity range of these dominant techniques is that the relative number of known exoplanets with wide-separation is small, thus our knowledge about such planets is still poor. 
This paucity of detections is especially marked for the population of planets beyond the snow line \citep{2004ApJ...616..567I, 2004ApJ...612L..73L, 2006ApJ...650L.139K}, usually defined as the distance from the host star in a stellar nebula at which water may condense into solid ice grains. 

Detecting exoplanets using gravitational microlensing was proposed by \citet{1964PhRv..133..835L} and \citet{1991ApJ...374L..37M} though it was described in notebooks and private communications as early as 1915 by Albert Einstein as a means to test his theoretical work regarding the deflection of light by mass. 
When a background source star is closely aligned with a foreground lens star, the gravity of the lens bends the light from the source star to create unresolved images of the source, yielding an apparent magnification of the source star brightness. 
The relative motion of the lens and source stars results in a lightcurve with brightness changing as a function of time. 
If the lens star has a planetary companion lying close to one of the source images, the gravity of the planet perturbs the image, producing an anomaly in the observed lightcurve.
Microlensing is sensitive to planets \citep{1996ApJ...472..660B} orbiting faint and/or distant stars and exhibits unique sensitivity to planets with orbital radii 1-6 AU, just outside the snow line, with masses down to that of Mercury.

The results of the statistical analysis of planets discovered from the MOA-II microlensing survey conducted during 2007 - 2012 period suggest that cold exo-Neptunes are the most common type of planets beyond the snow line \citep{2016ApJ...833..145S, 2018ApJ...869L..34S}.
These studies used the planet-host mass \textit{ratio}, the primary observable in all planetary microlensing events, to determine the exoplanet frequency. 
Other information is needed to obtain the actual masses of the system bodies from these measurements. 
A statistically robust sample of masses of planets beyond the snow line is important because it may permit more meaningful results to be drawn from a demographic understanding of exoplanets. 
In particular, such measurements hold the potential to provide a crucial calibration of planet formation theory. 

Obtaining such a robust sample of planet masses beyond the snow line is made particularly challenging due primarily to the difficulty in determining the masses of lens stars $M_{L}$ and the distances to the lens systems $D_{L}$ for general microlensing events. 
If we measure both the angular Einstein radius, $\theta_{\rm E}$ , and the microlensing parallax, $\pi_{\rm E}$, the mass and distance of the lens star may be uniquely determined \citep{1992ApJ...392..442G, 2008Sci...319..927G, 2011ApJ...741...22M}. 
The angular Einstein radius is defined as $\theta_{\rm E} \equiv (4GM / c^{2} D_{rel})^{1/2}$, where $M$ is the mass of the lens system, $D_{rel}^{-1} \equiv D_{L}^{-1} - D_{S}^{-1}$, and $D_{S}$ is the distance to the source star. 
The microlensing parallax, measured from two separated locations in the observer's plane, is defined as $\pi_{\rm E} \equiv {\rm AU} / (\theta_{\rm E} \, D_{rel})$. 
The angular Einstein radius may be directly measured for events in which the caustic crossing features in the lensing lightcurve are resolved. 
Here the term caustic refers to a closed locus of points in the magnification pattern created by the lensing system for which magnification formally approaches infinity. 
Using these techniques, we are able to obtain two different mass-distance relations from $\theta_{\rm E}$ and $\pi_{\rm E}$ thereby permitting us to resolve the degeneracy naturally arising in the mass ratio. 
The measurement of microlensing parallax is, however, relatively rare for events observed using only ground-based telescopes due to the short baselines between observatories, diurnal phase differences at observatory sites and deleterious observing conditions that may frustrate these time-critical measurements. 

Without the measurement of the microlensing parallax, one can still obtain an additional mass-distance relation from the lens flux by using a mass-luminosity relation.
Because the source stars are located in crowded stellar fields of the Galactic bulge, it is difficult to resolve the lens from nearby blended stars.
However, the contamination of the flux by blended stars can be greatly reduced if the flux is measured in high-resolution images obtained by using ground-based telescopes equipped with adaptive optics (AO) system or space telescopes \citep{2014ApJ...780...54B, 2015ApJ...808..170B, 2015ApJ...808..169B}.
However, even in high-resolution images, the ambient stars or a companion to the source or lens star can be blended with the lens or the lens and source \citep{2017AJ....154...59B, 2017AJ....154....3K}.
In the case where the lens and the source are not separated sufficiently to be resolved, we can measure the excess flux $f_{\rm excess}$, which is defined as $f_{excess} \equiv f_{target} - f_{S}$, where $f_{target}$ is the target flux obtained from high-resolution images, and $f_{\rm S}$ is the source flux obtained from lightcurve fitting.
\citet{2017AJ....154....3K} and \citet{2019Koshimoto} developed a method to evaluate the probability distributions of fluxes of the contaminants and the lens in the excess flux.
If the contamination probability is sufficiently small, the excess flux can be regarded as the lens flux, and we can uniquely measure the lens mass, $M_{L}$, and the distance to the lens system, $D_{L}$, from any combinations of $\theta_{\rm E}$, $\pi_{\rm E}$, and the lens flux.

In this paper, we report the discovery of a planet found from the analysis of the microlensing event OGLE-2015-BLG-1649.
We estimate the lens mass from the angular Einstein radius together with the excess flux measurement in the high-resolution images obtained from follow-up observations using the Subaru telescope with an AO system. 
We describe the observations by the microlensing survey and follow-up teams in Section \ref{sec-obs}.
Section \ref{sec-red} explains our data reduction procedure. 
The lightcurve modeling is described in Section \ref{sec-model}. 
Section \ref{sec-cmd} presents the source star radius estimate. 
In Section \ref{sec-ircs}, we present the Subaru observations and our analysis.
We evaluate the excess flux in Section \ref{sec-excess}.
Section \ref{sec-properties} shows the Bayesian analysis we used to estimate the posterior probability density distribution of the lens properties with the consideration of contamination probabilities to the lens flux. 
Finally, our discussion and conclusions are given in Section \ref{sec-discuss}. \\

\section{Observations}
\label{sec-obs}

On 2015 July 18, HJD$-2450000 \equiv {\rm HJD^{\prime}} = 7221$, the microlensing event OGLE-2015-BLG-1649 was discovered and alerted by the Optical Gravitational Lensing Experiment (OGLE; \citealp{2015AcA....65....1U}) Early Warning System (EWS). 
The source star of the event is located at $(\alpha, \delta)(2000) = (18^{h}04^{m}49^{s}_{\cdot}21,$ $-32^{\circ}37^{\prime}58^{\prime \prime}_{\cdot}90)$ which correspond to Galactic coordinates: $(l, b) = (-1^{\circ}_{\cdot}124, -5^{\circ}_{\cdot}422)$.
The Microlensing Observations in Astrophysics (MOA; \citealp{2001MNRAS.327..868B,2003ApJ...591..204S}) collaboration independently found the event, which was named as MOA-2015-BLG-404, and alerted the discovery on 2015 July 30. 

In the fourth phase of their survey, the OGLE collaboration is observing the Galactic bulge using the 1.3m Warsaw telescope located at Las Campanas Observatory in Chile \citep{2015AcA....65....1U}. 
The observations by OGLE were carried out in the $I$-band and occasionally in the $V$-band.
In the following analysis, we use the $V$-band data only for independent color measurement of the source.

The MOA collaboration is also conducting a microlensing exoplanet search towards the Galactic bulge, using the 1.8m MOA-II telescope at Mt. John Observatory (MJO) in New Zealand.
MOA conducts an efficient, nightly, high cadence survey using a wide 2.2 $\rm deg^{2}$ field of view (FOV) with a $\rm 10k \times 8k$ pixel mosaic CCD-camera, MOA-cam3 \citep{2008ExA....22...51S}. 
The observations by MOA were mainly with a custom broad $R + I$-band filter called MOA-Red and with a $V$-band filter called MOA-V. 
MOA also conducted follow-up observations by using the 61cm Boller \& Chivens (B\&C) telescope at MJO with simultaneous $g$-, $r$-, $i$-band imaging.

The MOA collaboration noticed an anomaly, which appeared to be a caustic entry, on 2015 August 11, ${\rm HJD^{\prime}} = 7246.1$, and issued an alert, prompting follow-up observations. 
The RoboNet collaboration \citep{2009AN....330....4T} conducted follow-up observations in the $I$-band using the Las Cumbres Observatory Global Telescope (LCOGT) Network 1.0m telescopes sited at CTIO/Chile, SAAO/South Africa, and Siding Spring/Australia \citep{2013PASP..125.1031B}.
In addition, the Microlensing Network for the Detection of Small Terrestrial Exoplanets (MiNDSTEp) conducted follow-up observations using the 1.54m Danish Telescope at the European Southern Observatory in La Silla, Chile \citep{2010AN....331..671D}. 
The MiNDSTEp data were collected using an EMCCD camera with a long-pass filter ${i_{\rm dk}}$ resembling an extended SDSS-i $+$ SDSS-z filter with a low-wavelength cut-off at 6500 {\AA} \citep{2015A&A...574A..54S, 2016A&A...589A..58E}. 

The lightcurves for these datasets are shown in Figure \ref{fig:lcurve}.
The number of data points are also shown in Table \ref{tab:err}. 

We conducted high-resolution imaging observations to constrain the lens flux 40 days after detection of the anomaly using the Subaru telescope. 
We describe the details of the Subaru observations and the analysis in Section \ref{sec-ircs}. \\

\section{Data Reduction}
\label{sec-red}

The OGLE and MOA data are reduced with the OGLE Difference Image Analysis (DIA) photometry pipeline \citep{2003AcA....53..291U} and MOA's implementation of a DIA pipeline \citep{2001MNRAS.327..868B}, respectively.
The RoboNet and MiNDSTEp data are reduced by using DanDIA \citep{2008MNRAS.386L..77B, 2013MNRAS.428.2275B}.
The DIA method has an advantage for the photometry of stars located in crowded fields such as the Galactic bulge field. 
It also produces better photometric lightcurves, because it is more efficient in dealing with the effect of blending compared to traditional PSF photometry. 

It is known that the nominal error-bars calculated by the pipelines are incorrectly estimated in such crowded stellar fields for various reasons. 
We employ a standard empirical error-bar normalization process \citep{2012ApJ...755..102Y} intended to estimate proper uncertainties for the lensing parameters in the lightcurve modeling. 
This process, described below, does not affect the lensing parameters. 
We renormalize the photometric uncertainty using the formula,
\begin{equation}
  \sigma^{\prime}_{i} = k \sqrt{\sigma^{2}_{i} + e^{2}_{min}},
\end{equation}
in which $\sigma^{\prime}_{i}$ is the renormalized uncertainty in magnitude, while $\sigma_{i}$ is uncertainty of the $i$th original data point obtained from DIA. 
The variables $k$ and $e_{min}$ are renormalizing parameters. 
For preliminary modeling, we search for the best-fit lensing parameters using $\sigma_{i}$.
We then construct a cumulative $\chi^{2}$ distribution as a function of lensing magnification. 
The $e_{min}$ value is chosen so that the slope of the distribution is 1. 
The $k$ value is chosen so that $\chi^{2}/dof \simeq 1$. 
In Table \ref{tab:err}, we list the so-derived error-bar renormalization parameters. \\

\section{Lightcurve Modeling}
\label{sec-model}

\begin{figure}[htbp]
  \begin{center}
    \includegraphics[scale=0.6,angle=-90]{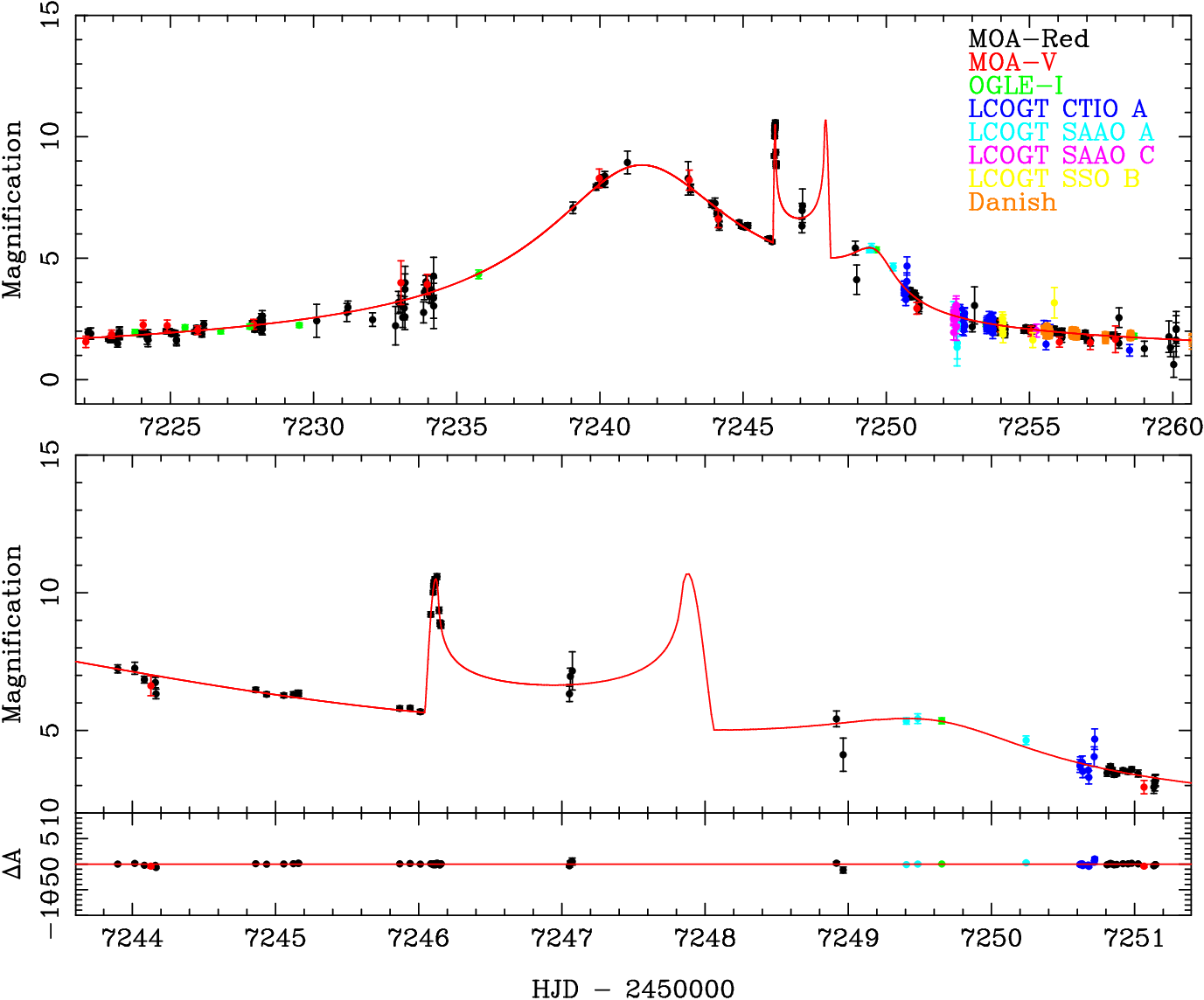}
    \caption{The lightcurve data of event OGLE-2015-BLG-1649 with the best-fit model. The best-fit model is indicated by the red line. Middle and bottom panels show the detail of planetary signal and the residual from the best model respectively. The data points taken by the B\&C telescope are not shown for display purposes but models have been fitted to these data, as well as the data from all other sources.}
  \label{fig:lcurve}
  \end{center}
\end{figure}

The caustic entry of this event is well observed by MOA. (See Figure \ref{fig:lcurve}.)
Unfortunately, while MOA was unable to observe the caustic exit, LCOGT data sample the critical caustic approach feature at ${\rm HJD^{\prime}} = 7249.5$.

There are five microlensing parameters for a point-source point-lens (PSPL) model: the time of the closest lens-source approach $t_{0}$, the Einstein radius crossing timescale $t_{\rm E}$, the impact parameter in units of the Einstein radius $u_{0}$, the source flux $f_{\rm S}$ and the blend flux $f_{\rm B}$. 
There are three more parameters for a point-source binary-lens model: the planet-host mass ratio $q$, the projected planet-host separation in units of the Einstein radius $s$ and the angle between the trajectory of the source and the planet-host axis $\alpha$. 
In case the finite size of the source is considered (finite source effect), we include a source size in units of the Einstein radius $\rho \equiv \theta_{\rm *}/\theta_{\rm E}$, where $\theta_{\rm *}$ is the angular source radius, and $\theta_{\rm E}$ is the angular Einstein radius of the lens. 
If microlensing parallax due to Earth's orbital motion is detected during the event, the north, $\pi_{\rm E,N}$ and east, $\pi_{\rm E,E}$, components of the microlensing parallax vector, $\bm{{\pi}_{\rm E}}$, are added. 
If effects from both finite source size and microlensing parallax are detected, we can uniquely determine the lens mass and the distance \citep{2011ApJ...741...22M, 2019AJ....157..215S}. 

We conduct lightcurve modeling using the Markov Chain Monte Carlo (MCMC) algorithm of \citet{2003ApJS..148..195V}. 
For the computation of finite-source magnification, we use the image-centered ray-shooting method \citep{1996ApJ...472..660B, 2010ApJ...716.1408B} implemented by \citet{2010ApJ...710.1641S}. 
The overall shape of the lensing lightcurve is parameterized by $(q,s,\alpha)$. 
We conduct a grid search for these parameters, starting from 9680 grid points, while we search for the remaining parameters using a downhill simplex method. 
Subsequently, we search for the best model among the leading 100 candidate models from the initial grid search by allowing all parameters to vary. 
In microlensing event OGLE-2015-BLG-1649, we detect a finite source effect and use linear limb-darkening coefficients for a Solar type star in the initial grid search and subsequent runs.
Once a candidate model is found, we further refine it with updated linear limb-darkening coefficients based on source color to obtain the best-fit model. 
The stellar effective temperature $T_{\rm eff}$, computed from the source color presented in Section \ref{sec-cmd}, is $T_{\rm eff} = 5777 \pm 571 ~ \rm K$ \citep{2009A&A...497..497G}. 
We assume $T_{\rm eff} \sim 5750 ~ \rm K$, a surface gravity of $\log g = 4.5$ ($g$ is in a unit of $\rm cm \, s^{-2}$), the microturbulent velocity as $v_{\rm t} = 1 \, \rm km \, s^{-1}$, and a metalicity of $\log [M/H] = 0$. 
We use the corresponding limb-darkening coefficients from the ATLAS stellar atmosphere models of \citet{2011A&A...529A..75C}, where the limb-darkening coefficients, $u_{\lambda}$, for these datasets are shown in Table \ref{tab:err}. 

\begin{figure}[htbp]
  \begin{center}
    \includegraphics[scale=0.6,angle=-90]{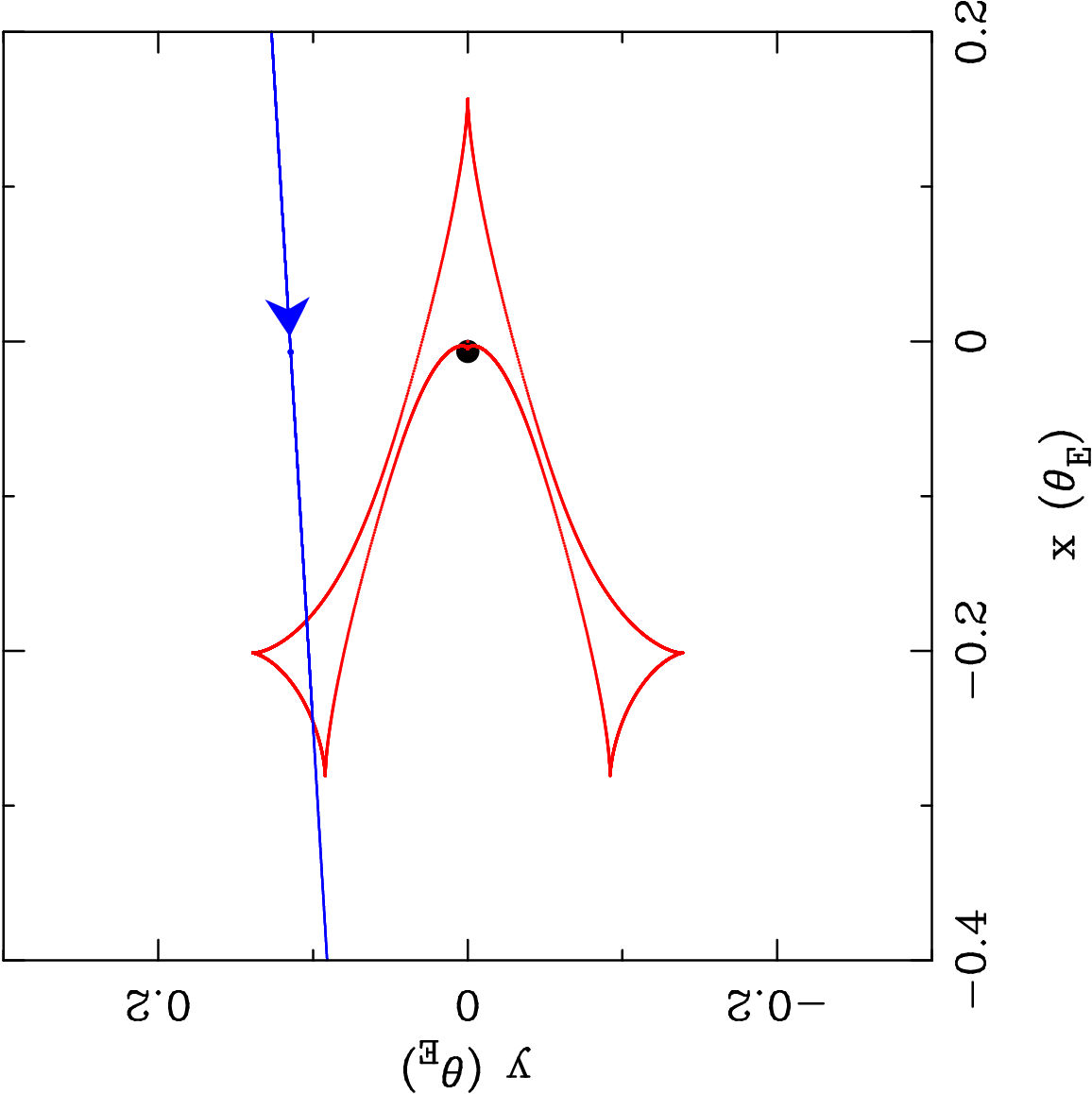}
    \caption{Caustic geometries for the best-fit model indicated by the red curve. The blue line shows the source trajectory with respect to the lens system. The blue circle indicates the source star size. The origin of the coordinate system corresponds to the barycenter of the lens system. The planet is located at (s, 0).}
  \label{fig:cau}
  \end{center}
\end{figure}

In Table \ref{tab:bestpara} and Figure \ref{fig:lcurve}, we present the lensing parameters and the model of the best-fit solution, respectively. 
In Figure \ref{fig:cau}, we also present the lens-system configuration in which the source trajectory with respect to the binary-lens caustic is shown. 
We find that the best-fit model has a planetary mass ratio of $q=(7.2 \pm 0.2) \times 10^{-3}$ and a projected separation $s=0.902 \pm 0.001$. 

Our analysis suggests that the proper motion of the source star causes it to cross the lensing system's caustic with the caustic entry and one crossing well sampled by the MOA data. 
Because the infinitesimally thin caustic effectively resolves the source star, inclusion of the finite source effect improves the fit by $\Delta \chi^{2} = 597.3$. 
In contrast, the inclusion of the microlensing parallax effect improves the fit by only $\Delta \chi^{2} = 5.4$, i.e., less than 2$\sigma$. 
We therefore adopt the best model, including the finite source effect while excluding parallax, in subsequent analysis. \\

\section{CMD and Source Radius}
\label{sec-cmd}

\begin{figure}[htbp]
  \begin{center}
    \includegraphics[scale=0.6,angle=-90]{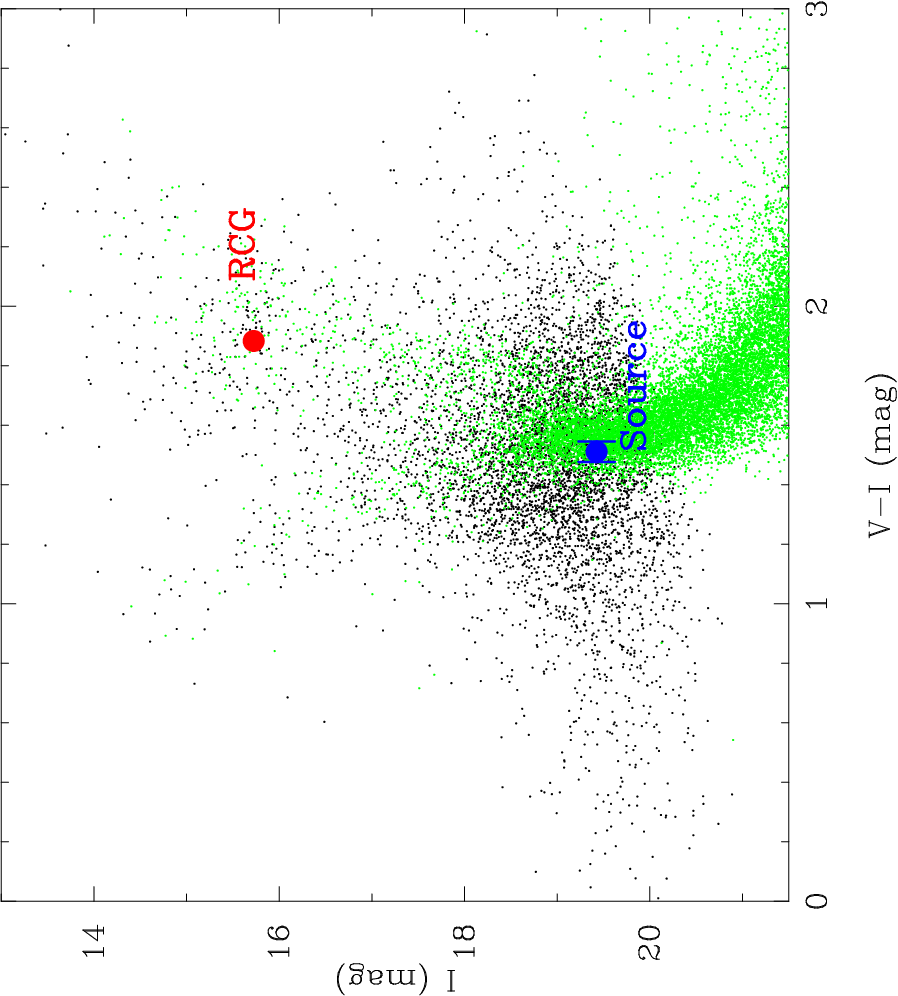}
    \caption{Color Magnitude Diagram (CMD) of OGLE-III stars within $2^{\prime}$ of OGLE-2015-BLG-1649 (black dots). The green dots show the HST CMD \citep{1998AJ....115.1946H}. The red point indicates the centroid of the red clump giants, and the blue point indicates the source color and magnitude.}
  \label{fig:cmd}
  \end{center}
\end{figure}

In this section, we estimate the angular Einstein radius $\theta_{\rm E}=\theta_{\rm *}/\rho$ from the combination of $\rho$ and $\theta_{\rm *}$, where the normalized source radius is measured from the lightcurve modeling and the angular source radius is estimated from the color and brightness of the source. 
We obtain the source color and magnitude by fitting the lightcurve to the MOA-Red band and MOA-V band data. 
Figure \ref{fig:cmd} shows the OGLE-III $(V-I,I)$ color-magnitude diagram (CMD) of stars within $2^{\prime}$ around the source \citep{2011AcA....61...83S}. 
It also shows the deep CMD of Baade's window observed by {\it HST} \citep{1998AJ....115.1946H}. 
The {\it HST} CMD is aligned to the ground-based CMD considering the distance, reddening and extinction to the OGLE-2015-BLG-1649 field by using Red Clump Giants (RCG) as standard candles \citep{2008ApJ...684..663B}. 
We convert the best fit MOA-Red and MOA-V source magnitude to the standard Cousins I and Johnson V magnitudes by cross-referencing stars in the MOA field with stars in the OGLE-III photometry map \citep{2011AcA....61...83S} within $2^{\prime}$ of the event. 
We find the source color and magnitude to be $(V-I,I)_{\rm S,OGLE} = (1.51 \pm 0.03, 19.43 \pm 0.02)$. 
We independently measure the source color by using OGLE-$I$ and $V$ lightcurves and we found $(V-I)_{\rm S,OGLE} = 1.52 \pm 0.09$ which is consistent with above value.
We use $(V-I)_{\rm S,OGLE} = 1.51 \pm 0.03$ in the following analysis.
The centroid of RCG color and magnitude in the CMD are $(V-I,I)_{\rm RCG} = (1.88 \pm 0.03, 15.73 \pm 0.06)$ as shown in Figure \ref{fig:cmd}. 
Comparing these values to the expected extinction-free RCG color and magnitude at this field of $(V-I,I)_{\rm RCG,0} = (1.06 \pm 0.07, 14.51 \pm 0.04)$ \citep{2013A&A...549A.147B, 2013ApJ...769...88N}, we get the reddening and extinction to the source of $(E(V-I),A_{I}) = (0.82 \pm 0.08, 1.22 \pm 0.07)$. 
Therefore, we estimate the extinction-free source color and magnitude as $(V-I,I)_{\rm S,0} = (0.69 \pm 0.08, 18.21 \pm 0.07)$.
By using the empirical formula, $\log(2 \theta_{\rm *}) = 0.5014 + 0.4197(V-I) - 0.2I$ \citep{2014AJ....147...47B, 2015ApJ...809...74F}, we estimate the angular source radius to be 
\begin{equation}
  \theta_{\rm *} = 0.70 \pm 0.06 \ {\rm \mu as}.
\end{equation}
From this $\theta_{\rm *}$ and other fitting parameters, we calculate the angular Einstein radius $\theta_{\rm E}$ and the lens-source relative proper motion $\mu_{\rm rel} = \theta_{\rm E} / t_{\rm E}$, as follows, 
\begin{eqnarray}
  \theta_{\rm E} & = & 0.57 \pm 0.06 \ {\rm mas} \\
  \mu_{\rm rel} & = & 7.14 \pm 0.67 \ {\rm mas \ yr^{-1}}.
\end{eqnarray}

\section{IRCS AO Images}
\label{sec-ircs}

We conducted high-resolution imaging follow-up observations of OGLE-2015-BLG-1649 using the Infrared Camera and Spectrograph (IRCS, \citealp{2000SPIE.4008.1056K}) with the adaptive optics system AO188 \citep{2010SPIE.7736E..0NH} mounted on the 8.2m Subaru Telescope on 2015 September 18 at 5:17-6:05 UT (${\rm HJD^{\prime}} = 7283.7$). 
We employed the high-resolution mode of IRCS, which delivers a pixel scale of 20.6 mas/pixel and a $21^{\prime \prime} \times 21^{\prime \prime}$ FOV. 
For AO correction, we use a bright star located close to the source star. 
We obtained 15 exposures in the $H$ and $K^{\prime}$-bands with 24-sec exposures with a five-point dithering and 15 $J$-band with 30-sec exposures with a five-point dithering. 
The AO-corrected seeing was $0^{\prime \prime}_{\cdot}37$, $0^{\prime \prime}_{\cdot}22$, and $0^{\prime \prime}_{\cdot}19$ for $J$, $H$, and $K^{\prime}$ images, respectively. 

Image reductions are carried out in a standard manner, including flat-fielding and sky-subtraction. 
We then combine all single-exposure images to form deep stacked images in each pass-band. 
The stacked images are further aligned with VISTA Variables in the Via Lactea (VVV, \citealp{2010NewA...15..433M}) images for astrometric calibration. 
We estimate the flux of OGLE-2015-BLG-1649 using aperture photometry. 
We conduct calibration in a \textit{photometric ladder} manner: we first calibrate the photometry of IRCS stacked images against the VVV data, and then scale to the Two Micron All Sky Survey (2MASS, \citealp{2006AJ....131.1163S}) photometric system. 
We find that the brightness of the event at the time of the AO observation is
\begin{eqnarray}
J_{target} & = & 18.467 \, \pm \, 0.189, \\
H_{target} & = & 17.870 \, \pm \, 0.217, \\
K^{\prime}_{target} & = & 17.667 \, \pm \, 0.127. 
\end{eqnarray} \\

\section{Excess Flux}
\label{sec-excess}

The measured angular Einstein radius provides a mass-distance relation, i.e., $M = (4G/c^2)^{-1} \theta_{\rm E}^{2} D_{rel}$. 
A second mass-distance relation may be estimated in the case where the lens flux is detected. 
If both relations can be measured the lens mass can be uniquely determined. 
High-resolution imaging with IRCS/Subaru gives us the combined flux from the lens, source and other blended stars. 
If we can obtain the source flux from lightcurve fitting, the total flux from the lens and blend can be calculated by subtracting the source brightness from the combined flux (equation 5-7). 
We do not have lightcurve data in $J$-, $H$- and $K$-bands.
Therefore, we derive the source magnitude in $H$-band as $H_{\rm S,0} = 17.57 \pm 0.12$ by converting $(V-I,I)_{\rm S,0} = (0.69 \pm 0.08, 18.21 \pm 0.07)$ with the color-color relation by \citet{1995ApJS..101..117K}.
We use the CMD of VVV to derive the extinction value in $H$-band, $A_{H}$.
We subsequently compare the centroid of RCG on CMD and the intrinsic position of RCG derived by \citet{2016MNRAS.456.2692N} resulting in an extinction value of $A_{H} = 0.41 \pm 0.12$.

The magnification at the time of Subaru observation is $A = 1.128$ according to the best-fit model.
The apparent $H$-band magnitude of the source at the time is expected to be $H_{S, AOtime} = 17.84 \pm 0.15$ in 2MASS system \citep{2010ApJ...711..731J, 2001AJ....121.2851C}.
This suggests that the $H$-band flux observed by Subaru mainly comes from the slightly magnified source.
We can place the $1 \sigma$ upper limit of the excess brightness of $H_{excess} > 19.11$. 
Using a similar process, we obtain $1 \sigma$ upper limits of excess brightness $J_{excess} > 20.18$ and $K_{excess} > 19.21$. \\

\section{Lens Properties through Bayesian Analysis}
\label{sec-properties}

\begin{figure}[htbp]
  \begin{center}
    \includegraphics[scale=0.6,angle=-90]{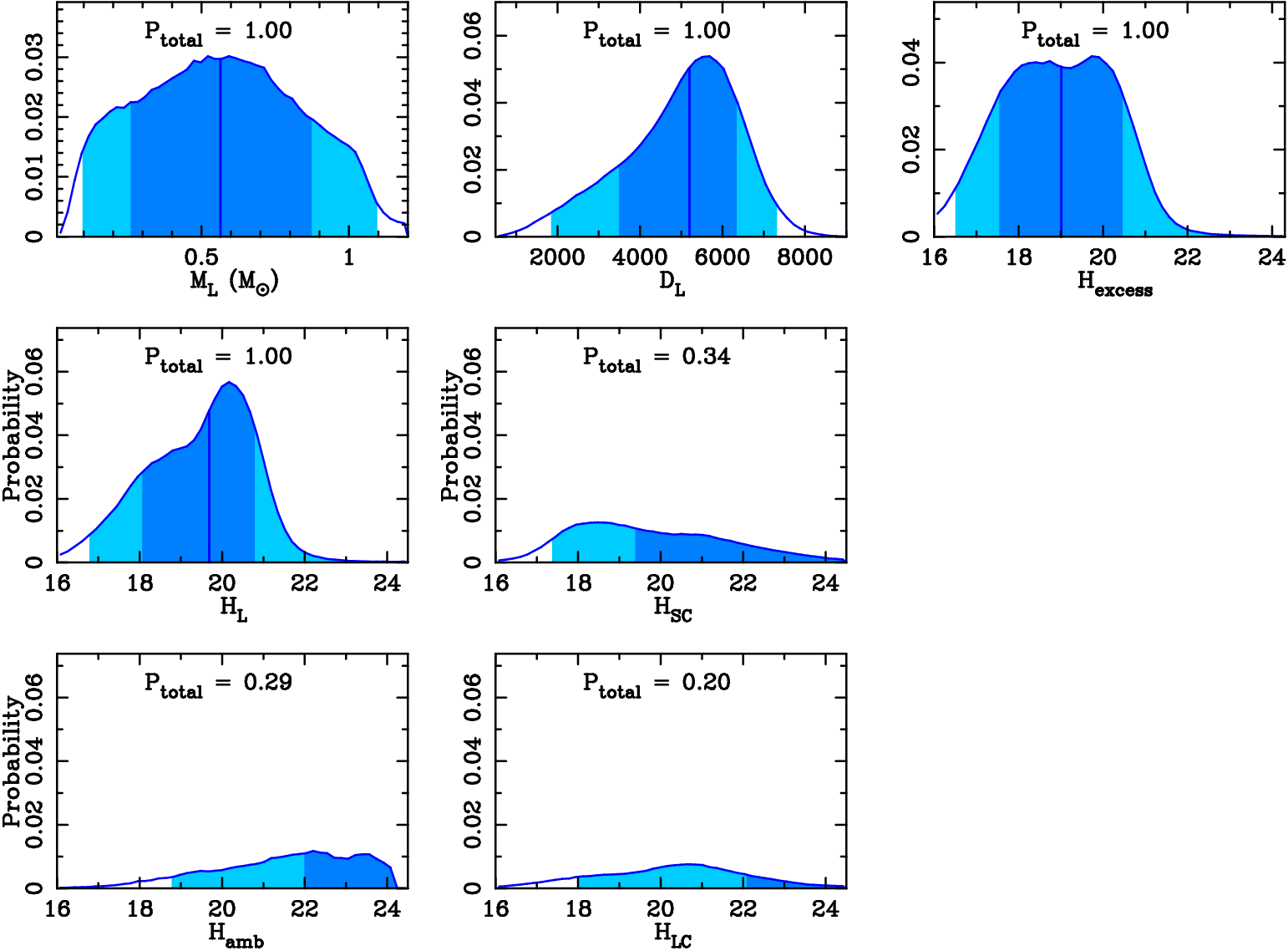}
    \caption{The posterior probability distributions of the lens mass $M_{L}$, the distance to the lens system $D_{L}$, total magnitude of contamination $H_{excess}$, magnitude of the lens star $H_{L}$, magnitude of source companion $H_{SC}$, magnitude of ambient star $H_{amb}$, and magnitude of lens companion $H_{LC}$. The dark and light blue regions indicate the 68$\%$ and 95$\%$ confidence intervals respectively. The vertical blue lines indicate the median values of each of these distributions. These distributions have not been constrained by the excess brightness limit.}
  \label{fig:pri}
  \end{center}
\end{figure}

To estimate the properties of the lens system, we consider the probability of possible sources of contamination (unrelated ambient stars, a companion to the source star, and a companion to the lens star) in the estimated excess H-band flux, $H_{excess}$ \citep{2014ApJ...780...54B, 2015ApJ...809...74F, 2017AJ....154....3K}. 
Following the method of \citet{2017AJ....154....3K}, we determine the posterior probability distributions of these sources for the origin of the excess flux. 
We use the Galactic model of \citet{1995ApJ...447...53H} as our prior distribution and the measured $\theta_{\rm E}$ and $t_{\rm E}$ to constrain the posterior probability distributions of lens parameters.
Figure \ref{fig:pri} shows the posterior probability distributions of the lens mass $M_{L}$, the distance to the lens system $D_{L}$, total magnitude of contamination $H_{excess}$, magnitude of the lens star $H_{L}$, magnitude of ambient star $H_{amb}$, magnitude of source companion $H_{SC}$, and magnitude of lens companion $H_{LC}$.

\begin{figure}[htbp]
  \begin{center}
    \includegraphics[scale=0.6,angle=-90]{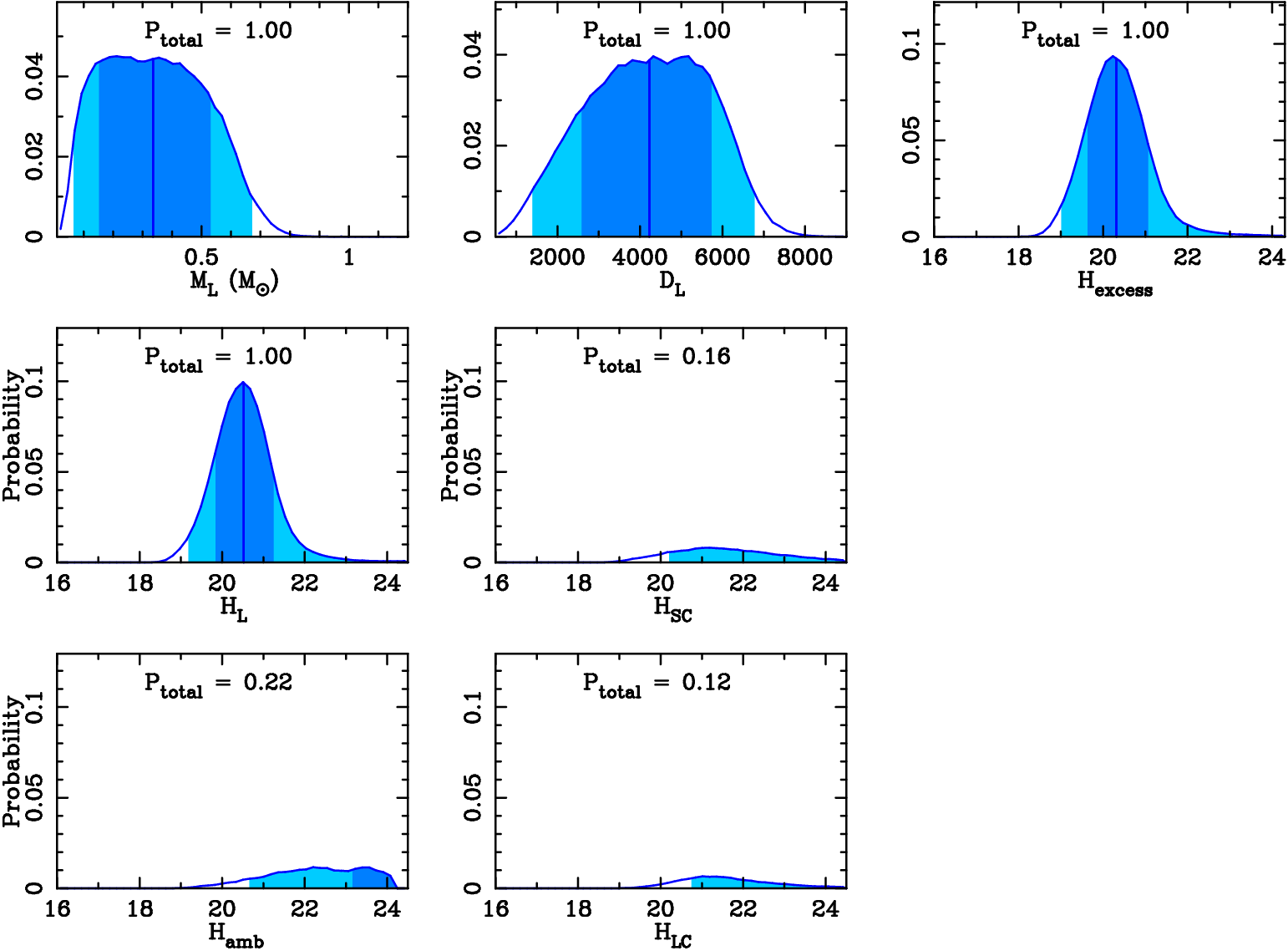}
    \caption{The posterior probability distributions narrowed by the additional constraint of the excess brightness limit.  The panel in the upper left suggests that the host star is almost certainly less massive than a G-dwarf.}
  \label{fig:post}
  \end{center}
\end{figure}

With the upper limit on lens brightness, we can make the posterior probability distribution much narrower.
We use the probability distributions in Figure \ref{fig:pri} to extract combinations of the parameters that satisfy the $1 \sigma$ upper limit of $H_{excess} > 19.11$. 
Figure \ref{fig:post} shows the posterior probability distributions with the additional constraint of the excess brightness limit. 
Table \ref{tab:lensprop} shows the median and $1\sigma$ range of $H_{L}$, $M_{L}$, $D_{L}$, the values of the planet mass $M_{p}$, the projected separation $a_{\perp}$ and the three-dimensional star-planet separation $a_{3d}$ for the posterior probability distribution with and without the excess brightness limit. 
The intrinsic orbital separation $a_{3d}$ is estimated assuming a uniform orientation of the planets, i.e., $a_{3d} = \sqrt{(3/2)} \times a_{\perp}$. 
More details can be found in \citet{2017AJ....154....3K}. 

While our Bayesian treatment of these data do not exclude the probability of a G-dwarf host (Figure \ref{fig:pri}, upper left), examination of the posterior distribution obtained with the constraint of the excess brightness limit (Figure \ref{fig:post}) allows us to assert that the host star is almost certainly less massive.
These results are consistent with similarly derived distributions for the $J$- and $K$-bands. 

The posterior distribution with excess brightness limit shows that the most likely lens brightness is $H_{L} = 20.52$.
Since the uncertainty of the source star magnitude in $H$-band is relatively large, we would have failed to detect the excess flux even if the seeing conditions were better during the Subaru observations.
Consequently, the lens and source stars must be spatially resolved to measure the $H$-band lens flux.
For this reason, this event is one of the high priority candidates for follow-up observations with high-resolution imaging because of the high relative proper motion and relative faintness of the source star. \\

\section{Discussion and Conclusion}
\label{sec-discuss}

We have here described the discovery of a planetary system, OGLE-2015-BLG-1649L, composed of a giant planet with $M_{p} = 2.5^{+1.5}_{-1.4} ~ M_{Jup}$ and an M or late K-dwarf host with $M_{L} = 0.34 \pm 0.19 ~ M_{\odot}$. 
Our analysis suggests that it is likely that the brightness values of possible sources of contamination are, in the aggregate, fainter than the brightness of the lens star. 
This suggests that the color-dependent centroid shift is likely to be caused by the lens itself. 
We estimate that the color-dependent centroid shift for this event will be $dx \sim 2.1$ mas in 2019 using the relation $dx = dt \times (f_{H} - f_{V}) \times \mu_{rel}$, in which $f_{H} = 0.09$ and $f_{V} = 0.01$ are the fraction of the lens $+$ source flux that is due to the lens in $H$- and $V$-band, respectively \citep{2007ApJ...660..781B, 2016ApJ...824..139H}. 
Although our Subaru AO observations were carried out when the source star was still magnified, we can yet obtain the source magnitude in $H$-band directly if additional Subaru observations are conducted in the near future. 
Considering the high relative proper motion, image elongation could be also measured with high-resolution observations in a few years' time \citep{2018AJ....156..289B}. 
For these reasons, this planetary microlensing event should be one of the highest priorities for future observation using a high-resolution instrument. 

To derive the cold planet frequency as a function of physical parameters, such as, host star mass, Galactocentric distance and planet mass function, it is manifestly desirable to use planet mass data that has been tightly constrained. 
IRCS AO observations permitted an estimate of an upper limit on the excess flux. 
This, in turn, provided a significantly tighter constraint on the lens flux than using the blending flux alone. 
While planetary parameters we have here estimated depend greatly on the prior distribution, our Bayesian analysis permits us exclude lens models in which the host star is a G-dwarf or a more massive star with relatively high credibility. 
In this study, we successfully demonstrated that we can reduce the uncertainty in host star mass by using an upper limit on the lens flux from AO images. 
Collecting AO imaged microlensing event data will be important for studying the planet mass function before the WFIRST \citep{2019ApJS..241....3P} era.

Finally, according to the standard core accretion model \citep{1972epcf.book.....S, 1985prpl.conf.1100H, 1993ARA&A..31..129L}, gas giant planets should seldom form around low-mass stars. 
By contrast, the disk instability model \citep{1997Sci...276.1836B} suggests no such restriction. 
Taken together with other gas giant/low-mass dwarf planetary systems that have been discovered (e.g., \citealp{2017AJ....154....3K}), OGLE-2015-BLG-1649Lb poses a challenge the former and appears to support the latter. \\

TS acknowledges the financial support from the JSPS, JSPS23103002, JSPS24253004 and JSPS26247023. 
The MOA project is supported by JSPS KAKENHI Grant Number JSPS24253004, JSPS26247023, JSPS23340064, JSPS15H00781, JP16H06287 and JP17H02871.
Work by N.K. is supported by JSPS KAKENHI Grant Number JP18J00897.
Work by Y.H. is supported by JSPS KAKENHI Grant Number JP17J02146.
The OGLE project has received funding from the National Science Centre, Poland, grant MAESTRO 2014/14/A/ST9/00121 to AU.
YT acknowledges the support of DFG priority program SPP 1992 "Exploring the Diversity of Extrasolar Planets" (WA 1047/11-1).

{}

\begin{table}[htb]
\caption{The data-sets for OGLE-2015-BLG-1649 and the error correction parameters.}
  \begin{center}
  \begin{tabular}{lccccc}
  \hline \hline
    Data-set & band & $k$ & $e_{min}$ & $u_{\lambda}$ & Number of data \\ \hline
    MOA & $R + I$ & 1.236825 & 0 & 0.53645 & 2668 \\
             & $V$ & 1.031747 & 0.038893 & 0.6556 & 184 \\
    OGLE & $I$ & 1.270605 & 0 & 0.4953 & 870 \\
    B\&C & $g$ & 0.728458 & 0 & 0.7276 & 125 \\
             & $r$ & 0.857322 & 0 & 0.6004 & 129 \\
             & $i$ & 0.760140 & 0 & 0.5152 & 125 \\
    LCOGT CTIO & $I$ & 1.031747 & 0 & 0.4953 & 56 \\
    LCOGT SAAO A & $I$ & 1.206830 & 0 & 0.4953 & 12 \\
    LCOGT SAAO C & $I$ & 1.128980 & 0 & 0.4953 & 15 \\
    LCOGT SSO B & $I$ & 1.454571 & 0 & 0.4953 & 10 \\
    Danish & ${i_{\rm dk}}$ & 0.491530 & 0 & 0.4543 & 86 \\
  \hline
  \label{tab:err}
  \end{tabular}
  \end{center}
\end{table}

\begin{table}[htb]
\caption{The best-fit parameters and $1\sigma$ errors.}
  \begin{center}
  \begin{tabular}{cccc}
  \hline \hline
    parameter & units & value & error ($1\sigma$) \\ \hline 
    $t_{0}$ & HJD - 2450000 & 7241.170 & 0.033 \\
    $t_{E}$ & days & 28.312 & 0.339 \\
    $u_{0}$ & $10^{-1}$ & 1.146 & 0.028 \\
    $q$    & $10^{-3}$ & 7.227 & 0.212 \\
    $s$    &        & 0.902 & 0.001 \\
    $\alpha$ & radians & 3.080 & 0.007 \\
    $\rho$ & $10^{-3}$ & 1.265 & 0.055 \\
    $\theta_{\rm *}$ & ${\rm \mu as}$ & 0.703 & 0.062 \\
    $\theta_{\rm E}$ & mas & 0.556 & 0.055 \\
    $\mu_{\rm rel}$  & ${\rm mas \ yr^{-1}}$ & 7.138 & 0.674 \\
    $d.o.f.$ &  & 4251 & \\
    $\chi^{2}$ &  & 4256.214 & \\
  \hline
  \end{tabular}
  \end{center}
  \label{tab:bestpara}
\end{table}

\begin{table}[htb]
\caption{Lens properties calculated from the posterior probability distribution with and without the Subaru AO data.}
  \begin{center}
  \begin{tabular}{cc|cc|cc}
  \hline \hline
  \multicolumn{2}{c}{} & \multicolumn{2}{|c}{w/o the Subaru data} & \multicolumn{2}{|c}{w/ the Subaru data}\\  \hline
    parameter    & units                 & median   & $1\sigma$ range & median & $1\sigma$ range \\ \hline
    $H_{L}$       & mag                  & 19.68 & 18.07 - 20.79       & 20.52      & 19.85 - 21.25 \\
    $M_{L}$       & $M_{\odot}$     & 0.56   & 0.26 - 0.87           & 0.34        & 0.15 - 0.53 \\
    $M_{p}$       & $M_{\rm Jup}$ & 4.27   & 1.96 - 6.61           & 2.54        & 1.15 - 4.02 \\
    $D_{L}$       & kpc                   & 5.20   & 3.50 - 6.34           & 4.23        & 2.59 - 5.74 \\
    $a_{\perp}$ & AU                    & 2.57   & 1.77 - 3.14           & 2.07        & 1.30 - 2.72 \\
    $a_{3d}$     & AU                    & 3.13   & 2.04 - 4.89           & 2.56        & 1.56 - 4.03 \\
  \hline
  \end{tabular}
  \end{center}
  \label{tab:lensprop}
\end{table}

\newpage


\begin{thebibliography}{}

\bibitem[Batista et al.(2015)]{2015ApJ...808..170B} Batista, V., Beaulieu, J.-P., Bennett, D.~P., et al.\ 2015, \apj, 808, 170
\bibitem[Batista et al.(2014)]{2014ApJ...780...54B} Batista, V., Beaulieu, J.-P., Gould, A., et al.\ 2014, \apj, 780, 54 
\bibitem[Bennett(2010)]{2010ApJ...716.1408B} Bennett, D.~P.\ 2010, \apj, 716, 1408
\bibitem[Bennett et al.(2007)]{2007ApJ...660..781B} Bennett, D.~P., Anderson, J., \& Gaudi, B.~S.\ 2007, \apj, 660, 781
\bibitem[Bennett et al.(2015)]{2015ApJ...808..169B} Bennett, D.~P., Bhattacharya, A., Anderson, J., et al.\ 2015, \apj, 808, 169
\bibitem[Bennett et al.(2008)]{2008ApJ...684..663B} Bennett, D.~P., Bond, I.~A., Udalski, A., et al.\ 2008, \apj, 684, 663-683
\bibitem[Bennett \& Rhie(1996)]{1996ApJ...472..660B} Bennett, D.~P., \& Rhie, S.~H.\ 1996, \apj, 472, 660
\bibitem[Bensby et al.(2013)]{2013A&A...549A.147B} Bensby, T., Yee, J.~C., Feltzing, S., et al.\ 2013, \aap, 549, A147
\bibitem[Bhattacharya et al.(2018)]{2018AJ....156..289B} Bhattacharya, A., Beaulieu, J.-P., Bennett, D.~P., et al.\ 2018, \aj, 156, 289
\bibitem[Bhattacharya et al.(2017)]{2017AJ....154...59B} Bhattacharya, A., Bennett, D.~P., Anderson, J., et al.\ 2017, \aj, 154, 59
\bibitem[Bond et al.(2001)]{2001MNRAS.327..868B} Bond, I.~A., Abe, F., Dodd, R.~J., et al.\ 2001, \mnras, 327, 868
\bibitem[Borucki et al.(2010)]{2010Sci...327..977B} Borucki, W.~J., Koch, D., Basri, G., et al.\ 2010, Science, 327, 977
\bibitem[Borucki et al.(2011)]{2011ApJ...736...19B} Borucki, W.~J., Koch, D.~G., Basri, G., et al.\ 2011, \apj, 736, 19
\bibitem[Boss(1997)]{1997Sci...276.1836B} Boss, A.~P.\ 1997, Science, 276, 1836
\bibitem[Boyajian et al.(2014)]{2014AJ....147...47B} Boyajian, T.~S., van Belle, G., \& von Braun, K.\ 2014, \aj, 147, 47
\bibitem[Bramich(2008)]{2008MNRAS.386L..77B} Bramich, D.~M.\ 2008, \mnras, 386, L77
\bibitem[Bramich et al.(2013)]{2013MNRAS.428.2275B} Bramich, D.~M., Horne, K., Albrow, M.~D., et al.\ 2013, \mnras, 428, 2275
\bibitem[Brown et al.(2013)]{2013PASP..125.1031B} Brown, T.~M., Baliber, N., Bianco, F.~B., et al.\ 2013, \pasp, 125, 1031
\bibitem[Butler et al.(2006)]{2006ApJ...646..505B} Butler, R.~P., Wright, J.~T., Marcy, G.~W., et al.\ 2006, \apj, 646, 505
\bibitem[Carpenter(2001)]{2001AJ....121.2851C} Carpenter, J.~M.\ 2001, \aj, 121, 2851
\bibitem[Claret \& Bloemen(2011)]{2011A&A...529A..75C} Claret, A., \& Bloemen, S.\ 2011, \aap, 529, A75
\bibitem[Dominik et al.(2010)]{2010AN....331..671D} Dominik, M., J{\o}rgensen, U.~G., Rattenbury, N.~J., et al.\ 2010, Astronomische Nachrichten, 331, 671
\bibitem[Evans et al.(2016)]{2016A&A...589A..58E} Evans, D.~F., Southworth, J., Maxted, P.~F.~L., et al.\ 2016, \aap, 589, A58
\bibitem[Fukui et al.(2015)]{2015ApJ...809...74F} Fukui, A., Gould, A., Sumi, T., et al.\ 2015, \apj, 809, 74
\bibitem[Gaudi et al.(2008)]{2008Sci...319..927G} Gaudi, B.~S., Bennett, D.~P., Udalski, A., et al.\ 2008, Science, 319, 927
\bibitem[Gonz{\'a}lez Hern{\'a}ndez \& Bonifacio(2009)]{2009A&A...497..497G} Gonz{\'a}lez Hern{\'a}ndez, J.~I., \& Bonifacio, P.\ 2009, \aap, 497, 497
\bibitem[Gould(1992)]{1992ApJ...392..442G} Gould, A.\ 1992, \apj, 392, 442
\bibitem[Han \& Gould(1995)]{1995ApJ...447...53H} Han, C., \& Gould, A.\ 1995, \apj, 447, 53
\bibitem[Hayano et al.(2010)]{2010SPIE.7736E..0NH} Hayano, Y., Takami, H., Oya, S., et al.\ 2010, \procspie, 7736, 77360N
\bibitem[Hayashi et al.(1985)]{1985prpl.conf.1100H} Hayashi, C., Nakazawa, K., \& Nakagawa, Y.\ 1985, Protostars and Planets II, 1100
\bibitem[Hirao et al.(2016)]{2016ApJ...824..139H} Hirao, Y., Udalski, A., Sumi, T., et al.\ 2016, \apj, 824, 139
\bibitem[Holtzman et al.(1998)]{1998AJ....115.1946H} Holtzman, J.~A., Watson, A.~M., Baum, W.~A., et al.\ 1998, \aj, 115, 1946
\bibitem[Ida \& Lin(2004)]{2004ApJ...616..567I} Ida, S., \& Lin, D.~N.~C.\ 2004, \apj, 616, 567
\bibitem[Janczak et al.(2010)]{2010ApJ...711..731J} Janczak, J., Fukui, A., Dong, S., et al.\ 2010, \apj, 711, 731
\bibitem[Kennedy et al.(2006)]{2006ApJ...650L.139K} Kennedy, G.~M., Kenyon, S.~J., \& Bromley, B.~C.\ 2006, \apjl, 650, L139
\bibitem[Kenyon \& Hartmann(1995)]{1995ApJS..101..117K} Kenyon, S.~J., \& Hartmann, L.\ 1995, \apjs, 101, 117
\bibitem[Kobayashi et al.(2000)]{2000SPIE.4008.1056K} Kobayashi, N., Tokunaga, A.~T., Terada, H., et al.\ 2000, \procspie, 4008, 1056
\bibitem[Koshimoto et al.(2017)]{2017AJ....154....3K} Koshimoto, N., Shvartzvald, Y., Bennett, D.~P., et al.\ 2017, \aj, 154, 3
\bibitem[Koshimoto et al.(2019, in preparation)]{2019Koshimoto} Koshimoto et al. in preparation \ 2018
\bibitem[Laughlin et al.(2004)]{2004ApJ...612L..73L} Laughlin, G., Bodenheimer, P., \& Adams, F.~C.\ 2004, \apjl, 612, L73
\bibitem[Liebes(1964)]{1964PhRv..133..835L} Liebes, S.\ 1964, Physical Review, 133, 835
\bibitem[Lissauer(1993)]{1993ARA&A..31..129L} Lissauer, J.~J.\ 1993, \araa, 31, 129
\bibitem[Mao \& Paczy{\'n}ski(1991)]{1991ApJ...374L..37M} Mao, S., \& Paczy{\'n}ski, B.\ 1991, \apjl, 374, L37
\bibitem[Mayor \& Queloz(1995)]{1995Natur.378..355M} Mayor, M., \& Queloz, D.\ 1995, \nat, 378, 355 
\bibitem[Minniti et al.(2010)]{2010NewA...15..433M} Minniti, D., Lucas, P.~W., Emerson, J.~P., et al.\ 2010, \na, 15, 433
\bibitem[Muraki et al.(2011)]{2011ApJ...741...22M} Muraki, Y., Han, C., Bennett, D.~P., et al.\ 2011, \apj, 741, 22
\bibitem[Nataf et al.(2016)]{2016MNRAS.456.2692N} Nataf, D.~M., Gonzalez, O.~A., Casagrande, L., et al.\ 2016, \mnras, 456, 2692
\bibitem[Nataf et al.(2013)]{2013ApJ...769...88N} Nataf, D.~M., Gould, A., Fouqu{\'e}, P., et al.\ 2013, \apj, 769, 88
\bibitem[Penny et al.(2019)]{2019ApJS..241....3P} Penny, M.~T., Gaudi, B.~S., Kerins, E., et al.\ 2019, \apjs, 241, 3
\bibitem[Safronov(1972)]{1972epcf.book.....S} Safronov, V.~S.\ 1972, Evolution of the protoplanetary cloud and formation of the earth and planets., by Safronov, V.~S..~Translated from Russian.~Jerusalem (Israel): Israel Program for Scientific Translations, Keter Publishing House, 212 p.,
\bibitem[Sako et al.(2008)]{2008ExA....22...51S} Sako, T., Sekiguchi, T., Sasaki, M., et al.\ 2008, Experimental Astronomy, 22, 51
\bibitem[Skottfelt et al.(2015)]{2015A&A...574A..54S} Skottfelt, J., Bramich, D.~M., Hundertmark, M., et al.\ 2015, \aap, 574, A54
\bibitem[Skrutskie et al.(2006)]{2006AJ....131.1163S} Skrutskie, M.~F., Cutri, R.~M., Stiening, R., et al.\ 2006, \aj, 131, 1163
\bibitem[Street et al.(2019)]{2019AJ....157..215S} Street, R.~A., Bachelet, E., Tsapras, Y., et al.\ 2019, \aj, 157, 215
\bibitem[Sumi et al.(2003)]{2003ApJ...591..204S} Sumi, T., Abe, F., Bond, I.~A., et al.\ 2003, \apj, 591, 204
\bibitem[Sumi et al.(2010)]{2010ApJ...710.1641S} Sumi, T., Bennett, D.~P., Bond, I.~A., et al.\ 2010, \apj, 710, 1641
\bibitem[Suzuki et al.(2018)]{2018ApJ...869L..34S} Suzuki, D., Bennett, D.~P., Ida, S., et al.\ 2018, \apjl, 869, L34
\bibitem[Suzuki et al.(2016)]{2016ApJ...833..145S} Suzuki, D., Bennett, D.~P., Sumi, T., et al.\ 2016, \apj, 833, 145
\bibitem[Szyma{\'n}ski et al.(2011)]{2011AcA....61...83S} Szyma{\'n}ski, M.~K., Udalski, A., Soszy{\'n}ski, I., et al.\ 2011, \actaa, 61, 83
\bibitem[Tsapras et al.(2009)]{2009AN....330....4T} Tsapras, Y., Street, R., Horne, K., et al.\ 2009, Astronomische Nachrichten, 330, 4
\bibitem[Udalski(2003)]{2003AcA....53..291U} Udalski, A.\ 2003, \actaa, 53, 291
\bibitem[Udalski et al.(2015)]{2015AcA....65....1U} Udalski, A., Szyma{\'n}ski, M.~K., \& Szyma{\'n}ski, G.\ 2015, \actaa, 65, 1
\bibitem[Verde et al.(2003)]{2003ApJS..148..195V} Verde, L., Peiris, H.~V., Spergel, D.~N., et al.\ 2003, \apjs, 148, 195
\bibitem[Yee et al.(2012)]{2012ApJ...755..102Y} Yee, J.~C., Shvartzvald, Y., Gal-Yam, A., et al.\ 2012, \apj, 755, 102
\bibitem[Yee et al.(2015)]{2015ApJ...802...76Y} Yee, J.~C., Udalski, A., Calchi Novati, S., et al.\ 2015, \apj, 802, 76

\end{thebibliography}
\end{document}